\input pipi.sty
\nopagenumbers
%\rightline\timestamp
\rightline{FTUAM 05-15}
\rightline{October,  2005}
%\rightline{hep-ph/??}

%---------
\bigskip
\hrule height .3mm
\vskip.6cm
\centerline{{\bigfib Some comments on calculations of the scalar radius of the pion}} 
\centerline{{\bigfib and the chiral constant $\bar{l}_4$\vphantom{\Bigg|}}}
\medskip
\centerrule{.7cm}
\vskip0.7truecm
\setbox9=\vbox{\hsize65mm {\noindent\fib F. J. 
Yndur\'ain} 
\vskip .1cm
\noindent{\addressfont Departamento de F\'{\i}sica Te\'orica, C-XI\hb
 Universidad Aut\'onoma de Madrid,\hb
 Canto Blanco,\hb
E-28049, Madrid, Spain.}\hb}
\smallskip
\centerline{\box9}
\bigskip

%--------
\setbox0=\vbox{\abstracttype{Abstract} The pion  scalar radius  
 is given by $\langle r^2_S\rangle=(6/\pi)\int_{4M^2_\pi}^\infty{\rm d}t\,\delta_S(t)/t^2$, 
with $\delta_S$  the phase of the scalar form factor. 
Below $\bar{K}K$ threshold, $\delta_S=\delta_0$,  $\delta_0$ being the isoscalar, 
S-wave $\pi\pi$ phase shift.  Between $\bar{K}K$ threshold and $t^{1/2}\sim 1.5\,{\rm GeV}$
 I  argued, in two previous letters,  that one can approximate   $\delta_S\sim\delta_0$, because 
inelasticity is small, compared with the errors. 
This gives $\langle r^2_S\rangle=0.75\pm0.07\,{\rm fm}^2$ and 
the value $\bar{l}_4=5.4\pm0.5$ for the one-loop chiral perturbation theory 
constant, compared with the values given by   Leutwyler and collaborators,  
 $\langle r^2_S\rangle=0.61\pm0.04\,{\rm fm}^2$ and 
 $\bar{l}_4=4.4\pm0.3$. At high energy, $t^{1/2}>1.5\,{\rm GeV}$, I remarked that 
the value of $\delta_S$ that follows from  perturbative QCD agrees with my interpolation and 
disagrees with that of  Leutwyler and collaborators. In a recent article, Caprini, Colangelo and
Leutwyler claim that my estimate of the asymptotic phase  $\delta_S$
is incorrect as it neglects higher twist contributions. Here I remark that, when correctly
calculated, higher twist contributions are likely negligible. I also show that  chiral perturbation
theory   gives  $\bar{l}_4=6.60\pm0.43$, compatible with my estimate but widely off the value
$\bar{l}_4=4.4\pm0.3$ of  Leutwyler and collaborators.}
\centerline{\box0}
\brochureendcover{Typeset with \physmatex}
\brochureb{\smallsc f. j.  yndur\'ain}{\smallsc the  
 scalar radius of the pion and the chiral constant $\;\scriptstyle{\bar{l}_4}\;$}{1}
\vskip1truecm

\booksection{1. Introduction.}

\noindent
 I will be concerned  
here mainly with four papers: 
they will be denoted by Y1,\ref{1} Y2,\ref{2} EY{\ref 3} and CCL.\ref{4}
$F_V(t)$, resp., $F_S(t)$ will be the vector  and scalar 
form factors of the pion; $\langle r^2_V\rangle$ and $\langle r^2_S\rangle$ 
the respective square radii. $\delta_V(t)$ will be the phase of  $F_V(t)$ and 
$\delta_S(t)$ the phase of $F_S(t)$. Finally,  $\delta_1(t)$, $\delta_0(t)$ 
are the P and S0 phase shifts
for $\pi\pi$ scattering.
 The pion  scalar radius  
 is given by 
$$\langle r^2_S\rangle=(6/\pi)\int_{4M^2_\pi}^\infty{\rm d}t\,\delta_S(t)/t^2.$$ 
For the vector radius one has a similar formula,
$$\langle r^2_V\rangle=(6/\pi)\int_{4M^2_\pi}^\infty{\rm d}t\,\delta_V(t)/t^2.$$
Below $\bar{K}K$ threshold, $\delta_S(t)=\delta_0(t)$ and  $\delta_V(t)=\delta_1(t)$.
In fact, this second equality probably holds as a good approximation 
up to $t\sim 2\;-\;3\,\gev^2$, where inelasticity begins to be important.

  Between $\bar{K}K$ threshold and
$t^{1/2}\sim 1.5\,{\rm GeV}$
 I  argued, in Y1 and Y2,  that one can approximate   $\delta_S\sim\delta_0$, because 
inelasticity is small compared with the errors. Using the experimental value 
for $\delta_0$,
this gives $\langle r^2_S\rangle=0.75\pm0.07\,{\rm fm}^2$ and 
the value $\bar{l}_4=5.4\pm0.5$ for the one-loop chiral perturbation theory 
(ch.p.t.) constant  $\bar{l}_4$, to be compared with the values given by   Leutwyler and
collaborators:\ref{4,5}  
 $\langle r^2_S\rangle=0.61\pm0.04\,{\rm fm}^2$ and 
 $\bar{l}_4=4.4\pm0.3$. 

I also argued in Y1 that at high energy, $t>2\,{\rm GeV}^2$, 
the value of $\delta_S$ that follows from  perturbative QCD agrees with my interpolation and 
disagrees with that of  Leutwyler and collaborators. In a recent article, Caprini, Colangelo and
Leutwyler (CCL\ref{4}) claim that my estimate of the asymptotic phase  $\delta_S$
is incorrect as it neglects higher twist contributions. 
Here I remark that, when correctly calculated, higher twist contributions are
very likely negligible. 
I also show that ch.p.t.  gives the number $\bar{l}_4=6.6\pm0.4$,
 compatible within reasonable two loop corrections with my estimate; but widely off the value
$\bar{l}_4=4.4\pm0.3$ of  Leutwyler and collaborators.

\booksection{2. The asymptotic  $F_V(t)$, $F_S(t)$}

\noindent
It is generally assumed that one can approximate, at 
high energy,  $F_V(t)$ by its  asymptotic form at
 leading twist (l.t.) the 
expression for which,   neglecting quark and pion masses, has been 
known for a long time:
$$F_V(t)\simeqsub_{t\to\infty}F_V^{\rm (l.t.)}(t)\simeqsub_{t\to\infty}\dfrac{12\pi
C_Ff^2_\pi\alpha_s(t)}{-t}=
\dfrac{12\pi}{33-2n_f}\;\dfrac{12\pi C_Ff^2_\pi}{-t\log(-t/\lambdav^2)}.
\equn{(1)}$$
(For the definition of leading twist and twist three contributions 
to the pion form factor, see ref.~3).

In 1983, Espriu and I calculated the twist three (t3) correction to this,\ref{3} obtaining an 
infrared divergent result,
$$F_V^{\rm
(t3)}(t)\simeqsub_{t\to\infty}\dfrac{f^2_\pi m^2_\pi}{(\hat{m}_u+\hat{m}_d)^2}\;
\dfrac{C}{-t}[\log(-t/M_0)]^{3+2d_m};
\equn{(2)}$$
the $\hat{m}$ are the invariant quark masses, $\bar{m}(t)=
\tfrac{1}{2}\hat{m}\log^{d_m} t/\lambdav^2$, and 
$M_0$ is a cutoff mass (that may be taken equal to $\lambdav$ or $m_\pi$;  
it is irrelevant which, as the difference is subleading).
We were not able to calculate the numerical constant $C$; for the
other terms, cf. EY, especially
the expressions for $\hat{B}_0(\nu)$ and the first formula in the 
second column of p. 189 in ref.~3.  So, the full form factor would
be
$$\eqalign{
F_V(t)\simeqsub_{t\to\infty}&\,F_V^{\rm (l.t.)}(t)+F_V^{\rm (t3)}(t)
\simeqsub_{t\to\infty}\dfrac{12\pi
C_Ff^2_\pi\alpha_s(t)}{-t}
+\dfrac{f^2_\pi m^2_\pi}{(\hat{m}_u+\hat{m}_d)^2}\;
\dfrac{C}{-t}[\log(-t/M_0)]^{3+2d_m}\cr
\simeqsub_{t\to\infty}&\,\dfrac{f^2_\pi m^2_\pi}{(\hat{m}_u+\hat{m}_d)^2}\;
\dfrac{C}{-t}[\log(-t/M_0)]^{3+2d_m},\cr}
\equn{(3)}$$
the last because, at large $t$, the (t3) contribution dominates the 
(l.t.) one. Note that the (t3) contribution also dominates the (l.t.) one 
in the chiral limit, for all $t$.

When the paper was in the preprint stage, we received a letter from Brodsky and 
Lepage, pointing out that infrared divergent terms could not be 
treated as we had done; one should instead introduce a cutoff in $\xi$ 
(the fraction of momentum of the quarks inside the pion) 
and treat the region near the
endpoints as in refs.~6. If one does this, the  twist three contribution becomes
$$F_V^{\rm (t3)}(t)\simeqsub_{t\to\infty}\dfrac{{\rm Const.}}{-t^{3/2}}
\equn{(4)}$$
instead of (2), and the (l.t.) piece dominates: one recovers (1), for all of $F_V$.

In EY\ref{3} we were not fully convinced by
 the Brodsky--Lepage arguments [although we quoted 
them, including the result in Eq. (4) in the present paper], because a fit to data 
with (1) was much worse than a fit with (2) or (3), as we remarked in 
EY. Thus, 
in EY we concluded that 
``one really cannot calculate the pion form factor in perturbative QCD."

However, some time later, Isgur and Llewellyn~Smith\ref{7} 
 considered the partonic wave function  of
the pion $\phi(\xi,t)$, for finite $t$.
 They concluded that the
asymptotic limit is reached  slowly and that,
 when reasonable $\phi(\xi,t)$ for finite $t$ are used in the (l.t.) expression 
for $F_V$, the corresponding (l.t.) expression fits data
well, even at subasymptotic energies. 
[An important remark is that the {\sl phase} of 
$F_V$ is {\sl not} affected by the change in the wave function piece, which is real].
 Because of this, in ref.~8  I used  
  much milder words: I merely wrote that 
``There are, unfortunately, a number of
snags" with the behavior (1), 
and proceeded to quote the EY, Brodsky--Lepage and Isgur and Llewellyn~Smith\ref{7}
 remarks, without drawing 
a definite conclusion.

For $F_S(t)$ the situation is very similar. To (l.t.) and leading order in 
perturbation theory, 
$$
F_S^{\rm (l.t.)}(t)\simeqsub_{t\to\infty}
\dfrac{576\pi^2}{33-2n_f}\;
\dfrac{f^2_\pi (\hat{m}_u+\hat{m}_d)^2}{-t(\log(-t/\lambdav^2)^{2d_m}},
\equn{(5)}$$
as shown in Y2, Eq.~(6.4).
But, according to CCL, the (t3) contribution is\fnote{I am
 not sure of the correctness of the expression  given in (6) for the 
(t3) piece; in EY we could not calculate analytically the twist-three wave function 
of the pion, $\phi_P$, hence we could  not get the
constant in Eq.~(3), either. In fact we did not get constant asymptotic $\phi_P$
 but found it growing like a power of $\log t$.
I have the impression that CCL have forgotten that one has to diagonalize 
the anomalous dimension matrix, i.e., that you have to calculate what 
 we call $T_{nm}$ in EY, which enters in the final result: 
see the detailed discussion in refs.~3,~8.
 However, since the conclusions are not
altered  by this, we will forget this problem here.}
$$F_S^{\rm (t3)}(t)\simeqsub_{t\to\infty}
\dfrac{48\pi^2f^2_\pi M^2_\pi}{33-2n_f}\;\dfrac{\log(-t/M_0)}{-t}.
\equn{(6)}$$
In this case, (6) would dominate over (5), both at large $t$ 
and in the chiral limit, and thus, in particular at large $t$, we would have
$$\eqalign{
F_S(t)\simeqsub_{t\to\infty}&\,F_S^{\rm (l.t.)}(t)
+F_S^{\rm (t3)}(t)
\simeqsub_{t\to\infty}\dfrac{576\pi^2}{33-2n_f}\;
\dfrac{f^2_\pi (\hat{m}_u+\hat{m}_d)^2}{-t(\log(-t/\lambdav^2)^{2d_m}}+\dfrac{48\pi^2f^2_\pi
M^2_\pi}{33-2n_f}\;\dfrac{\log(-t/M_0)}{-t}\cr
\simeqsub_{t\to\infty}&\,
\dfrac{48\pi^2f^2_\pi
M^2_\pi}{33-2n_f}\;\dfrac{\log(-t/M_0)}{-t}.\cr
}
\equn{(7)}$$

However, if one  takes into account the Brodsky--Lepage remarks,
 the (t3) part should 
be calculated \`a la Brodsky--Lepage--Drell, would behave 
as $t^{-3/2}$, and hence would be negligible: 
one would  recover the behaviour (5) given in Y1, for all of $F_S$.
Note that the situation is identical for 
$F_V$ and $F_S$; the difference between 
(l.t.) and (t3) contribution is due to the different pion wave functions involved, axial 
$\phi$ or
pseudoscalar, $\phi_P$, and these are the same for $F_V$ and $F_S$.

\booksection{3. The asymptotic  $\delta_V(t)$, $\delta_S(t)$}

\noindent
The 
phases of the form factors depend crucially on what happens to the (t3) contributions. 
If we believe the naive calculations in EY, or CCL, for the $F^{\rm (t3)}$, i.e., 
that the behaviour of the form factors is given by (3) and (6), then, 
$$\eqalign{
\delta_V(t)\simeqsub_{t\to\infty}&\,\pi\left\{1-(3+2d_m)\dfrac{1}{\log t/M_0}\right\}\quad
[{\rm with}\; (3)],\cr 
\delta_S(t)\simeqsub_{t\to\infty}&\,\pi\left\{1-\dfrac{1}{\log t/M_0}\right\}\quad\quad
[{\rm with}\; (6)]:\cr
}
\equn{(8)}$$
both phases approach the asymptotic value $\pi$ from {\sl below}.
However, if we accept the Brodsky--Lepage arguments, one gets instead 
dominant (l.t.) and thus
$$\eqalign{
\delta_V(t)\simeqsub_{t\to\infty}&\,\pi\left\{1+\dfrac{1}{\log t/M_0}\right\}\quad
[{\rm with}\; (1)],\cr 
\delta_S(t)\simeqsub_{t\to\infty}&\,\pi\left\{1+2d_m\dfrac{1}{\log t/M_0}\right\}\quad\quad
[{\rm with}\; (5)],\cr
}
\equn{(9)}$$
and the value $\pi$ is now approached from above.

The last results, (9), were what I used in Y1, Y2. In fact, and because 
(as one can easily imagine) I was aware of the (t3) threat, I 
started by checking, in Y1, 
the behaviour (9) for $\delta_V(t)$: here one has an {\sl independent} 
evaluation of  $\langle r^2_V\rangle$, from the {\sl experimental}
pion form factor,\ref{9}
  with which to compare. Looking at pp.~102,~103 in Y1, and correcting for an 
error there (which  alters nothing numerically; see the Erratum in Y1), 
we find the following result:
$$\langle r^2_V\rangle=\cases{0.434\quad {\rm fm}^2:\quad{\rm with}\;(9);
\cr0.303\quad {\rm fm}^2:\quad{\rm with}\;(8).\cr }
\equn{(10)}$$
We have assumed the asymptotic region to begin at 
$t=3\,\gev^2$ and taken $\delta_V(t)\simeq \delta_1(t)$ below this.
Experimentally,\ref{9} $\langle r^2_V\rangle=0.432\pm0.006$ 
(the error here includes electromagnetic effects). 
 The agreement of what one gets with (9) with
experiment is very good;  it only depends  a little 
on the values of $\lambdav$ or the matching point,
that I took to be 1.7~\gev, provided we keep them reasonable.
However, and this is more important,  if  using the
naive (t3) result,  given in Eq.~(8), 
the value obtained is a disaster: below experiment by 30\%. 
It is because of this that I took seriously, in Y1 (and in Y2) the 
Brodsky--Lepage arguments and hence the (l.t.) evaluation of the 
{\sl phase} of the form factors, Eq.~(9).  

Nevertheless, I concede that the situation is not clear. While there exist  
a number of indications favouring the results of Y1, Y2 
for $\langle r^2_{S}\rangle$, one cannot  
definitely exclude the results of the evaluations of the Leutwyler group, 
as already remarked in Y2.
One may therefore summarize the situation as follows:
First of all, the experimental data for $\pi\pi$ scattering 
in the important region $2m_K\leq t^{1/2}\leq 1.5\,\gev$ are ambiguous. 
Different analyses produce mutually incompatible results, so the value for
  $\langle r^2_{{\rm S}}\rangle$ depends on which data one uses. 
If choosing the central value in the fit of Hyams et al.\ref{10} 
one gets $\langle r^2_{{\rm S}}\rangle=0.61\pm0.04\,{\rm fm}^2$ and hence
$\bar{l}_4=4.4\pm0.3$ as in ref.~5.
However, and as discussed in Y2, if one uses different sets of data 
(notably, taking into account information from $\pi\pi\to\bar{K}K$ scattering) 
one finds a larger value,
 $\langle r^2_{{\rm S},\pi}\rangle=0.75\pm0.07\,{\rm fm}^2$, leading to 
$\bar{l}_4=5.4\pm0.5$. 
It could be possible, in principle, to discriminate between both 
solutions by using the asymptotic value for the phase of 
$F_S(t)$ that follows from perturbative QCD; 
the result, however, is infrared divergent and thus, as happens
 for the twist three contribution to the electromagnetic
 form factor of the pion, a cut-off becomes necessary and the 
result is ambiguous. Therefore, one should probably accept 
a conservative estimate, covering both evaluations, 
$$\bar{l}_4=5\pm1.
\equn{(11)}$$

\booksection{4. $\bar{l}_4$ in chiral perturbation theory}

\noindent
But this is not all. 
As I remarked in Y1,  the number of low energy observables in $\pi\pi$
scattering, to one loop in ch.p.t., is such that one can determine the 
value of $\bar{l}_4$ by fitting them, even if leaving  $\bar{l}_4$  free.

In the present note I elaborate on this. 
To do so, I will use the values for scattering lengths and
effective range  parameters  given in Table~I. 
In  units with the charged pion mass $M_\pi=1$, one has  

\medskip
{\bigskip
\setbox0=\vbox{\petit
\setbox1=\vbox{ \offinterlineskip\hrule
\halign{
&\vrule#&\strut\hfil\ #\ \hfil&\vrule#&\strut\hfil\ #\ \hfil&
\vrule#&\strut\hfil\ #\ \hfil&\vrule#&\strut\hfil\ #\ \hfil\cr
 height2mm&\omit&&\omit&&\omit&&\omit&\cr 
&\hfil \hfil&&\hfil ABT  \hfil&&
ACGL [CGL]  &  &\hfil Exp. [PY] \hfil& \cr
 height1mm&\omit&&\omit&&\omit&&\omit&\cr
\noalign{\hrule} 
height1mm&\omit&&\omit&&\omit&&\omit&\cr
&$a_0^{(0)}$&
&\vphantom{\Big|}$0.219\pm0.005$&
&$0.240\pm0.060\;[0.220\pm0.005]$&
&
$0.230\pm0.010$& \cr 
\noalign{\hrule}
height1mm&\omit&&\omit&&\omit&&\omit&\cr
&$a_0^{(2)}$&&\vphantom{\Big|}$-0.042\pm0.001$&&$-0.036\pm0.013\;[-0.0444\pm0.0010]$&&
$-0.052\pm0.012$& \cr 
\noalign{\hrule}
height1mm&\omit&&\omit&&\omit&&\omit&\cr
&$b_0^{(0)}$&&\vphantom{\Big|}$0.279\pm0.011$&&$0.260\pm0.02\;[0.276\pm0.006]$&&
$0.268\pm0.011$& \cr 
\noalign{\hrule}
height1mm&\omit&&\omit&&\omit&&\omit&\cr
&$b_0^{(2)}$&&\vphantom{\Big|}$-0.076\pm0.002$&&$-0.079\pm0.005\;[-0.080\pm0.001]$&&
$-0.085\pm0.011$& \cr 
\noalign{\hrule} 
height1mm&\omit&&\omit&&\omit&&\omit&\cr
&$10^3\times a_1$&&\vphantom{\Big|}$37.8\pm2.1$&&$37\pm2\;[37.9\pm0.5]$&&
$38.4\pm0.8$&\cr
\noalign{\hrule}
height1mm&\omit&&\omit&&\omit&&\omit&\cr
&$10^3\times b_1$&&\vphantom{\Big|}$5.9\pm1.2$&&$5.4\pm0.4\;[5.67\pm0.13]$&&
$4.75\pm0.16$& \cr 
\noalign{\hrule} 
height1mm&\omit&&\omit&&\omit&&\omit&\cr
&$10^4\times a_2^{(0)}$&&\vphantom{\Big|}$22\pm4$&&$17\pm1\;[17.5\pm0.3]$&&
$18.70\pm0.41$&\cr
\noalign{\hrule} 
height1mm&\omit&&\omit&&\omit&&\omit&\cr
&$10^4\times a_2^{(2)}$&&\vphantom{\Big|}$2.9\pm1.0$&&$1.8\pm0.8\;[1.70\pm0.13]$&&
$2.78\pm0.37$&\cr
\noalign{\hrule}}
\vskip.05cm}
\centerline{\box1}
\bigskip

{\noindent{\petit CGL, ACGL:  ref.~12, 
who apply dispersion relations together with experimental information at 
high energy and ch.p.t. (in CGL). ABT:  ref.~13; these authors 
 use models and chiral $SU(3)$. 
 Exp. [PY]: the experimental numbers as obtained in ref.~14
by fitting experimental data. The notation for the $a$s and $b$s is as in ref.~11.}}
\medskip
\centerline{\sc Table~I}
\smallskip
\centerrule{5truecm}}
\box0
}
\medskip

These $a$s and $b$s can be described in terms of only four chiral coupling 
constants, $\bar{l}_i,\;i=1,\,2,\,3,\,4$. In fact we will  include in the fits the 
estimate of Gasser and Leutwyler\ref{11} for $\bar{l}_3$ which gives 
$$\bar{l}_3=3\pm2.5.
\equn{(12)}$$
 This has little influence  the other parameters, but  imposing 
(12)  avoids unphysical values for  $\bar{l}_3$; otherwise the 
fits tend to give large negative values for this parameter, with very large errors.

We turn to the determination of 
$\bar{l}_4$ from ch.p.t.
We will use the one loop expressions for $a$s and $b$s in terms of $\bar{l}_i$ given 
in ref.~11,
$$\eqalign{
a_0^{(0)}=&\,\dfrac{7M_\pi}{32\pi f^2_\pi}\left\{1+\dfrac{5M^2_\pi}{84\pi^2f^2_\pi}
\left[\bar{l}_1+2\bar{l}_2-\tfrac{3}{8}\bar{l}_3+
\tfrac{21}{10}\bar{l}_4+\tfrac{21}{8}\right]\right\},\cr
a_0^{(2)}=&\,\dfrac{M_\pi}{16\pi f^2_\pi}\left\{1-\dfrac{M^2_\pi}{12\pi^2f^2_\pi}
\left[\bar{l}_1+2\bar{l}_2+\tfrac{3}{8}\right]+
\dfrac{M^2_\pi}{32\pi^2f^2_\pi}\left[\bar{l}_3+4\bar{l}_4\right]\right\},\cr
a_1^{(1)}=&\,\dfrac{1}{24\pi M_\pi  f^2_\pi}\left\{1-\dfrac{M^2_\pi}{12\pi^2f^2_\pi}
\left[\bar{l}_1-\bar{l}_2+\tfrac{65}{48}\right]+
\dfrac{M^2_\pi}{8\pi^2f^2_\pi}\bar{l}_4\right\};\cr
}
$$
$$\eqalign{
b_0^{(0)}=&\,\dfrac{1}{4\pi M_\pi f^2_\pi}\left\{1+\dfrac{M^2_\pi}{12\pi^2f^2_\pi}
\left[2\bar{l}_1+3\bar{l}_2-\tfrac{13}{16}\right]+
\dfrac{M^2_\pi}{8\pi^2f^2_\pi}\bar{l}_4\right\},\cr
b_0^{(2)}=&\,\dfrac{1}{8\pi M_\pi f^2_\pi}\left\{1-\dfrac{M^2_\pi}{12\pi^2f^2_\pi}
\left[\bar{l}_1+3\bar{l}_2-\tfrac{5}{16}\right]+
\dfrac{M^2_\pi}{8\pi^2f^2_\pi}\bar{l}_4\right\},\cr
b_1^{(1)}=&\,\dfrac{1}{288\pi^3 M_\pi f^4_\pi}
\left\{-\bar{l}_1+\bar{l}_2-\tfrac{97}{120}\right\}.\cr
}
$$
(for the D waves, see below) and we will consider
several fitting strategies:

\medskip
\noindent i) We include in the fit the constraints (11,~12), as well as the experimental 
numbers (PY) for $a$s and $b$s given in Table~I. Then, we get 
a $\chidof=8.8/(10-4)$ and the numbers
$$\matrix{
\bar{l}_1=&\,-1.30\pm0.23,&\quad \bar{l}_2=6.04\pm0.07,\quad
\bar{l}_3=&\,2.2\pm2.5,&\quad \bar{l}_4=6.32\pm0.61.\cr
}
\equn{(13a)}$$

The clear excess of the \chidof\ over unity 
indicates that, at the level of accuracy of ref.~14,
 the two loop effects are not negligible.

\medskip
\noindent ii) In fact,  the errors in (13a) are purely nominal, 
as the fit depends on the way we have treated the higher order corrections.
For example, if in the one loop expressions for $a_2^{(I)}$, $b_1^{(1)}$,
$$\eqalign{
a_2^{(0)}=&\,\dfrac{1}{1440\pi^3 M_\pi f^4_\pi}
\left\{\bar{l}_1+4\bar{l}_2-\tfrac{53}{8}\right\},\cr
a_2^{(2)}=&\,\dfrac{1}{1440\pi^3 M_\pi f^4_\pi}
\left\{\bar{l}_1+\bar{l}_2-\tfrac{103}{40}\right\};\cr
b_1^{(1)}=&\,\dfrac{1}{288\pi^3 M_\pi f^4_\pi}
\left\{-\bar{l}_1+\bar{l}_2-\tfrac{97}{120}\right\}.
}
$$ 
we replace the physical $f_\pi$ by its value in the chiral limit, $f$,
$$
f_\pi=f\left\{1+\dfrac{M^2_\pi}{16\pi^2f^2_\pi}\bar{l}_4\right\}, 
$$  which is allowed 
since the difference is of higher order in ch.p.t., 
we find a $\chidof=9.7/(10-4)$ and the central values 
$$\bar{l}_1=-0.33,\quad\bar{l}_2=4.56,\quad\bar{l}_3=2.1,\quad\bar{l}_4=6.93:
\equn{(13b)}$$
$\bar{l}_1$ and $\bar{l}_2$ are well outside the error bars given in (13a).

\medskip
\noindent iii) We can get safer error estimates, which take into account at least some
of the uncertainty due to higher order effects, by averaging (13a) 
and (13b), weighted with the 
respective \chidof, and 
enlarging the error by including, as an extra error, the difference between 
the result of this operation and (13a). 
In this way we find an  estimate that covers both (13a) and (13b):
$$\matrix{
\quad \bar{l}_1=-0.84\pm0.51,&\quad \bar{l}_2=5.34\pm0.70,\vphantom{\big|}
\quad\bar{l}_3=2.2\pm2.5,\quad \bar{l}_4=6.61\pm0.68.\cr
}
\equn{(13c)}$$

\medskip
\noindent iv) The two loop corrections to $\pi\pi$ scattering
 have also been calculated.\ref{15} 
It turns out that the scattering amplitude depends on six combinations of one and two loop 
coupling constants. Including two loop effects 
it is possible to give a virtually perfect fit 
to  experimental $\pi\pi$ scattering, up to the F wave; but, unlike for the one loop case 
where eight scattering lengths and effective range parameters are 
fitted with only two $\bar{l}$s, we now have six parameters for 
14 or 16 observables (see below), so the agreement, 
although better,  is  less impressive than before.
We find
$$\matrix{
\quad \bar{l}_1=-0.80\pm0.21,&\quad \bar{l}_2=5.5\pm0.35,
\quad\bar{l}_3=1.9\pm2.4,\quad \bar{l}_4=6.60\pm0.43.\cr
}
\equn{(13d)}$$
The numbers in (13d)  have been obtained as follows: 
one fits  the two loop expressions for scattering lengths and effective range 
parameters to the data deduced from experiment in ref.~14, 
 for all waves up to the F-wave.
One also includes in the fit  
 the constraint given by Eqs.~(13c) 
for the $\bar{l}_i$. We then find a $\chidof=3.8/(14-10)$, if {\sl not} 
including the F-wave in the fit, and a $\chidof=7.9/(16-10)$
if including $a_3^{(1)}$ and $b_3^{(1)}$ in the fit.
Averaging them and including their spread as an extra error
(that can be attributed to {\sl three} loop effects) we get (13d).

This result in (13d), $\bar{l}_4=6.60\pm0.43$, which is compatible with all others, is
what we consider the more
reliable one.

\medskip
\noindent v) The previous determinations include the constraint (11) 
in the fits. We could also fit leaving 
$\bar{l}_4$ completely free. In this case we find
$$\matrix{
\quad \bar{l}_1=-1.27\pm0.21,&\quad \bar{l}_2=6.04\pm0.06,
\quad\bar{l}_3=2.4\pm2.5,\quad \bar{l}_4=7.1\pm0.8,\cr
}
\equn{(14)}$$
with only the experimental errors in Table~I, which is compatible with (13). 
We would find  results similar 
to (13b,~c,~d) if including estimated 
higher order effects: the errors for $\bar{l}_4$ are slightly larger than before, 
and the central value is also only slightly larger.
This confirms the result in (13d) for $\bar{l}_4$.

\medskip
\noindent vi) Finally, we may also replace in the fits
 the ``Experimental" PY data with those of 
ACGL or ABT in Table~I. 
With only the errors there, {\sl and leaving 
$\bar{l}_4$ free}, we get
$$\matrix{
\quad \bar{l}_1=-2.3\pm0.1,&\quad \bar{l}_2=6.20\pm0.05,
\quad\bar{l}_3=1.0\pm2.0,\quad \bar{l}_4=6.6\pm0.3,\qquad\hbox{[CGL]}\vphantom{\Big|}\cr
\quad \bar{l}_1=-2.3\pm0.6,&\quad \bar{l}_2=6.0\pm0.3,
\quad\bar{l}_3=2.9\pm2.5,\quad \bar{l}_4=5.4\pm1.4,\qquad\hbox{[ACGL]}\vphantom{\Big|}\cr
\quad \bar{l}_1=-1.32\pm0.55,&\quad \bar{l}_2=6.84\pm0.28,
\quad\bar{l}_3=1.1\pm1.9,\quad \bar{l}_4=6.71\pm0.48.\qquad\hbox{[ABT]}\cr
}
\equn{(15)}$$

The numbers that follow for $\bar{l}_4$,  for {\bf all} the fits, 
 support the estimate in Y1, Y2,\fnote{The difference, 
$1.2\pm0.7$, i.e., a $20\pm16\%$ of the value following from the 
scalar form factor, is easily attributed to 
two loop effects in $\langle r^2_S\rangle$.}
$\bar{l}_4=5.4\pm0.5$,   
against that of Donoghue, Gasser and Leutwyler and  
Ananthanarayan et al.,\ref{5} $\bar{l}_4=4.4\pm0.3$.
It is particularly remarkable that the CGL evaluation of low energy parameters, 
which produces the parameters in the corresponding column in Table~I, 
and  
which I used for the evaluation CGL in Eq.~(15), was made imposing for  $\bar{l}_4$ 
the value of the Leutwyler group,  $\bar{l}_4=4.4\pm0.3$: 
requiring consistency of the low energy parameters, however, pushes 
 $\bar{l}_4$ to a value compatible with mine in Y1, Y2 and definitely incompatible 
with the   $\bar{l}_4=4.4\pm0.3$ input. 
In fact, and apart from the fit using the 
ACGL\ref{12} low energy parameters, 
whose large errors for $\bar{l}_4$ preclude drawing any clear conclusion 
from it,
 all other fits agree, in central value and error, 
with (13d). To get  results 
for the value of  $\bar{l}_4$ from the scalar radius of refs.~4,~5 consistent
with what one gets from $\pi\pi$ scattering,  one would require that the  
two loop corrections to the results of the Leutwyler group for $\langle r^2_S\rangle$ be 
of  50\% or more. 

\booksection{5. Reggeistics}

\noindent
Some comments are also made in ref.~{4} about Reggeistics. 
 I will only note here the surprising fact that, although a wealth
 of experimental data exist 
for high energy $\pi\pi$ cross sections,\ref{16} 
which were included in the fits in refs.~17,
Caprini, Colangelo and Leutwyler\ref{4} do not use them to derive what they call ``our
estimate" (see Fig.~2 in ref.~4) and they 
do not  compare what follows from said estimate with the $\pi\pi$ data.
The interested reader may find more detailed 
discussions of the
Reggeistics of the Leutwyler group in refs.~14,~17,~18.

\booksection{Acknowledgements}

\noindent 
I am grateful to J.~R.~Pel\'aez for very useful suggestions about the presentation 
in this note.

\booksection{References}

\noindent
1.- (Y1): Yndur\'ain, F. J.,  {\sl Phys.  Letters}, {\bf B578}, 99 and (E)
{\bf 586}, 439 (2004).\hb
2.- (Y2): Yndur\'ain, F. J.,  {\sl Phys. Letters}, {\bf B612}, 245 (2005).\hb
3.- (EY): Espriu, D., and Yndur\'ain, F. J. (1983). {\sl Phys. Letters}, {\bf B132}, 187.
See also
Efremov, A. V., and Radyushin, A. V., {\sl Riv. Nouvo Cimento} {\bf 3}, No. 2 
 (1980), where the relevance of the twist three contribution was first noticed.
\hb
4.- (CCL): Caprini, I, Colangelo, G., and Leutwyler, H., hep-ph/0509266 
 (2005).\hb
5.- {Donoghue, J. F., Gasser, J.,  and Leutwyler, H., {\sl Nucl. Phys.}, 
{\bf B343}, 341 (1990), slightly 
improved in Colangelo,~G., Gasser,~J.,  and Leutwyler,~H., {\sl Nucl. Phys.}, 
{\bf B603}, 125 (2001); 
 Ananthanarayan, B., et al., {\sl Phys. Lett.} {\bf B602}, 218 (2004).}\hb
6.- Brodsky, S. J., and Lepage, G. P., 
{\sl Phys. Rev.} {\bf D22}, 2157 (1980); 
Drell, S. D., and Yan, T. M.  {\sl Ann. Phys.} (N.Y.) {\bf 66}, 578 (1971); 
West, G.,  {\sl Phys. Rev.  Letters} {\bf 24}, 1206 (1970).
See also    Drell, S. D.,  and Hearn, A. C.,
{\sl  Phys. Rev. Lett.} {\bf 16}, 908 (1966).\hb
7.- Isgur, N., and Llewellyn Smith, C. H., {\sl Nucl. Phys.} {\bf B317}, 526
 (1989).\hb
8.- Yndur\'ain, F. J., {\sl The theory of quark and gluon interactions}, 
Springer, 1999 (3rd edition).\hb
9.- de~Troc\'oniz, J.~F., and Yndur\'ain, F.~J., 
 {\sl Phys. Rev.}  {\bf D65}, 093001
 (2002) and   {\sl Phys. Rev.} {\bf D71}, 073008 
(2005).\hb
10.- Hyams, B., et al., {\sl Nucl. Phys.} {\bf B100}, 205 (1975).\hb
11.- Gasser, J., and Leutwyler, H.,  {\sl Ann. Phys.} (N.Y.) {\bf 158}, 142 (1984).\hb
12.- (ACGL): Ananthanarayan, B., et al., {\sl Phys. Rep.}, {\bf 353}, 207,  (2001);
(CGL): Colangelo, G., Gasser, J.,  and Leutwyler, H.,
 {\sl Nucl. Phys.} {\bf B603},  125, (2001).\hb
13.- Amor\'os,~G., Bijnens, J., and Talavera, P., {\sl Nucl. Phys.} {\bf B585} 293 (2000)
and (E),  {\bf B598}, 665 (2001).\hb
14.- Pel\'aez, J. R., and Yndur\'ain,~F.~J., {\sl Phys. Rev.} {\bf D71}, 074016 (2005).\hb
15.- Knecht, M., et al., {\sl Nucl. Phys.} {\bf B457} 523 (1995);
Bijnens, J., et al., {\sl Nucl. Phys.} {\bf B508} 263 (1997) and (E), 
{\bf B517}, 639 (1998). See also ref.~13.\hb
16.- Biswas, N. N., et al., {\sl Phys. Rev. Letters}, 
{\bf 18}, 273 (1967) [$\pi^-\pi^-$, $\pi^+\pi^-$ and $\pi^0\pi^-$];
 Cohen, D. et al., {\sl Phys. Rev.}
{\bf D7}, 661  (1973) [$\pi^-\pi^-$];
 Robertson, W. J.,
Walker, W. D., and Davis, J. L., {\sl Phys. Rev.} {\bf D7}, 2554  (1973)  [$\pi^+\pi^-$]; 
Hoogland, W., et al.  {\sl Nucl. Phys.}, {\bf B126}, 109 (1977) [$\pi^-\pi^-$];
Hanlon, J., et al,  {\sl Phys. Rev. Letters}, 
{\bf 37}, 967 (1976) [$\pi^+\pi^-$]; Abramowicz, H., et al. {\sl Nucl. Phys.}, 
{\bf B166}, 62 (1980) [$\pi^+\pi^-$]. These  references cover the region
between  1.35 and 16 \gev, and agree within errors in the regions where they overlap 
(with the exception of $\pi^-\pi^-$ below 2.3 \gev).\hb
17.-  Pel\'aez, J.~R., and Yndur\'ain, F. J., {\sl Phys.
Rev.} {\bf D69}, 11400 (2004); Pel\'aez, J.~R. hep-ph/0510005 (2005).\hb
18.-  Pel\'aez, J. R., and Yndur\'ain, F.
J., {\sl Phys. Rev.} {\bf D68}, 074005 (2003); Pel\'aez, J. R., and Yndur\'ain, F. J.,
hep-ph/0412320  (Published on the Proc. of the Meeting ``Quark Confinement and 
the Hadron Spectrum", 
Villasimius, Sardinia, September 2004;  Pel\'aez, J. R., and Yndur\'ain, F. J., 
hep-ph/0510216, to appear in the Proc. of the 2005 Montpelier conference.\hb

\bye